# Controlling sharpness, SNR and SAR for 3D FSE at 7T by end-to-end learning

Peter Dawood*[1,2], Martin Blaimer[3], Jürgen Herrler[4], Patrick Liebig[4], Simon Weinmüller[1], Shaihan Malik[5], Peter M. Jakob[2], Moritz Zaiss[1,6]

[1]Institute of Neuroradiology, University Hospital Erlangen, Friedrich-Alexander-Universität Erlangen-Nürnberg, Erlangen, Germany

[2]Experimental Physics 5, University of Würzburg, Würzburg, Germany

[3]Magnetic Resonance and X-ray Imaging Department, Fraunhofer Institute for Integrated Circuits IIS, Division Development Center X-Ray Technology, Würzburg, Germany

[4]Siemens Healthcare, Erlangen, Germany

[5]Department of Biomedical Engineering, King's College, London, United Kingdom

[6]Department Artificial Intelligence in Biomedical Engineering, Friedrich-Alexander-Universität Erlangen-Nürnberg, Erlangen, Germany

*Correspondance to: Peter Dawood

    Institute of Neuroradiology

    University Hospital Erlangen

    peter.dawood@uk-erlangen.de

Abstract

Purpose

To non-heuristically identify dedicated variable flip angle (VFA) schemes optimized for the point-spread function (PSF) and signal-to-noise ratio (SNR) of multiple tissues in 3D FSE sequences with very long echo trains at 7T.

Methods

The proposed optimization considers predefined SAR constraints and target contrast using an end-to-end learning framework. The cost function integrates components for contrast fidelity (SNR) and a penalty term to minimize image blurring (PSF) for multiple tissues. By adjusting the weights of PSF/SNR cost-function components, PSF- and SNR-optimized VFAs were derived and tested in vivo using both the open-source Pulseq standard on two volunteers as well as vendor protocols on a 7T MRI system with parallel transmit extension on three volunteers.

Results

PSF-optimized VFAs resulted in significantly reduced image blurring compared to standard VFAs for T2w while maintaining contrast fidelity. Small white and gray matter structures, as well as blood vessels, are more visible with PSF-optimized VFAs. Quantitative analysis shows that the optimized VFA yields 50% less deviation from a sinc-like reference PSF than the standard VFA. The SNR-optimized VFAs yielded images with significantly improved SNR in a white and gray matter region relative to standard (81.2±18.4 vs. 41.2±11.5, respectively) as trade-off for elevated image blurring.

Conclusion

This study demonstrates the potential of end-to-end learning frameworks to optimize VFA schemes in very long echo trains for 3D FSE acquisition at 7T in terms of PSF and SNR. It paves the way for fast and flexible adjustment of the trade-off between PSF and SNR for 3D FSE.

## 1. Introduction

The fast-spin echo (FSE, also known as turbo-spin echo, TSE, or rapid acquisition with relaxation enhancement, RARE) (1) is an echo train sequence that is the backbone of modern, time-efficient MR applications. It is robust to magnetic field inhomogeneities and provides images with relevant contrast and high signal-to-noise ratio (SNR). Thus, it is widely used in clinical examinations. However, a major limitation is the high radiofrequency (RF) power deposition caused by the 180° flip angles, which often limits FSE examinations at high field strengths. The specific absorption rate (SAR) characterizes the RF power deposition and depends on the square of the main magnetic field and the square of the flip angles (2). SAR limits exist to prevent patient heating, however, even FSE examinations at 3 Tesla can exceed these limits. A state-of-the-art approach for SAR-reduced 2D multi-slice imaging is the Transitions Between Pseudo Steady States (TRAPS) method (3). In TRAPS, the use of reduced variable flip angle (VFA) schemes significantly reduces the SAR while maintaining the desired contrast and high signal strength.

Another major limitation is the reduced time efficiency of 3D acquisitions. In 2D FSE, multiple slices can be excited and acquired in an interleaved manner during each repetition time (TR). In 3D imaging, however, the entire volume is excited, and therefore, slice interleaving is not possible. Instead, long echo trains are preferred to acquire as much data as possible during each TR. As a drawback, the transverse magnetization (i.e., the MR signal) decays during the acquisition in the transient state within a TR, resulting in k-space data that are inherently modulated by a multi-component relaxation decay. The modulation is mainly determined by the respective tissue T2 relaxation parameters, and is represented by the modular transfer function (MTF). In a Fast Fourier transform reconstruction, a non-uniform MTF results in pronounced T2-dependent image blurring in the phase- and partition encoding direction, which is a major cause of image degradation and corrupts valuable image information.

To counteract T2 signal decay, VFA schemes are typically assigned to the refocusing pulses in the echo train (4)(5). By using dedicated VFA schemes, the magnetization at each echo is no longer refocused solely in the transverse plane. Instead, a significant portion of the magnetization is stored along the longitudinal axis. The subsequent use of this stored magnetization results in pseudo-steady states and enhanced signal amplitudes at relatively long echo times. This results in a more uniform MTF. For such approaches, the FA trains are optimized by performing a signal simulation to produce the desired signal response.

However, the desired signal responses are chosen heuristically. Although there are guidelines for the design of the VFA schemes, the user must define parameters and shapes for the ramp flip angles, for example (3)(4)(6). For T2-weighted imaging, the desired signal response is divided into three parts, with an asymptotic approach to a pseudo-steady state, a constant signal evolution until the k-space center is reached, and a linear flip angle increase until a maximum flip angle is reached (4)(5). While this standard approach produces clinical T2 contrast, the resulting VFAs may not be the optimal choice in terms of maximizing SNR or minimizing image blurring, which are important goals for specific imaging scenarios. Furthermore, the VFA design is typically limited to one target tissue and does not consider sampling schemes or spatial multi-tissue image content. Thus, while VFA schemes for 3D FSE are well-established approaches, potential improvements in terms of flexible, robust tailoring of VFA schemes to maximize SNR or minimize image blurring remain to be elucidated.

For ultrahigh field (UHF) MRI, the tailoring of dedicated VFA schemes is a particularly important aspect. While UHF MRI at 7 Tesla offers several advantages, such as higher baseline SNR, higher spectral resolution and, in many cases, better contrast between different tissues or between tissue and vessels, both SAR limits and image blurring due to faster T2 decay are serious issues. This is because the SAR depends on the square of the main magnetic field strength. In addition, reduced T2 relaxation parameters cause more signal loss for very long echo trains. Furthermore, MRI at 7T is strongly affected by inhomogeneous transmit B1 fields resulting in signal variations throughout the FOV, which can be mitigated with parallel transmit (pTx).

Compared to previous approaches (7)(8)(9)(10)(11), we suggest an end-to-end optimization framework for the flexible, non-heuristic identification of VFA schemes designed to maximize SNR or minimize image blurring of 3D FSE sequences with very long echo trains at 7T. Our framework considers multiple tissues, including their spatial distributions and sampling characteristics for clinical contrasts. VFA schemes are automatically identified to minimize image blurring or maximize SNR. An open-source 3D FSE sequence via the Pulseq standard is employed to test optimized VFA schemes and to quantify the reduction in image blurring and the improvement in SNR. Furthermore, the compatibility with clinical vendor sequences at 7T with pTx extension is shown and 3D acquisitions with very long echo trains lengths up to 220 are presented.

2. Methods

First, the optimization framework utilized to obtain VFA schemes tailored to PSF- or SNR optimization is described. Afterwards, the experimental evaluation is elaborated using an open-source, vendor-agnostic sequence development platform for imaging. It is then described how to quantify the benefits of PSF- and SNR optimization. Finally, the VFA schemes were teste using clinical vendor sequences.

### Standard VFA scheme

The gold standard VFA scheme (4)(5) uses a prospective extended phase graph (EPG) (12) algorithm to calculate the flip angle plan based on a target signal. The target signal is divided into three parts. First, an asymptotic approximation to a constant target signal level is computed. The target signal is kept constant until the k-space center is reached, and then a linear flip angle increase is applied until a maximum flip angle is reached. The VFA scheme that produces the defined signal response for a given T1 and T2 is then calculated. In this work, T1=1500 ms and T2=50 ms were chosen to optimize for WM.

2.1. Optimization framework

Although there exist other frameworks for VFA optimization such as MR-zero (10), we focus on a simplified model to reduce the computational complexity.

The proposed framework (Figure 1) employs a constrained minimization algorithm that treats FAs within an optimal control problem in which the total loss function is a combination of multiple losses:

$$\alpha^* = \arg\min_{\alpha} w \cdot L_{PSF}(\alpha) + (1-w) \cdot L_1(c_{TRG} \cdot I_{TRG}, I_{FSE}(\alpha)) + R(\alpha) \quad (1)$$

$$s.t. \ rSAR(\alpha) = c_{SAR}$$

Here, $\alpha = (\alpha_1, \ldots, \alpha_{ETL})$ is the FA vector (ETL: echo-train length), $I_{FSE}(\alpha)$ is the simulated FSE image for a given $\alpha$, $I_{TRG}$ is a target image scaled with parameter $c_{TRG} \in [0,1]$, $L_{PSF}$ is a PSF penalty, $L_1$ is a contrast/SNR loss and $R$ is a regularization term. The parameter $w$ controls the balance between PSF- and SNR optimization. The parameter $c_{SAR} \in [0,1]$ limits the relative SAR given by (2)

$$rSAR(\alpha) = \frac{1}{ETL}\sum_{i=1}^{ETL}\left(\frac{\alpha_i}{180°}\right)^2 \quad (2)$$

This parameter is used to provide the basis for a fair comparison of several sequences with different VFAs.

The contrast/SNR loss is chosen as the $L_1$ error of the simulated FSE image $I_{FSE}(\alpha)$ to the scaled target image $I_{TRG}$, i.e.

$$L_1(c_{TRG} \cdot I_{TRG}, I_{FSE}(\alpha)) = \frac{1}{N}|I_{FSE} - c_{TRG} \cdot I_{TRG}| \quad (3)$$

where $N$ is the total voxel number. The target $I_{TRG}$ is a simulated spin-echo image (mono-exponential decay) with predefined contrast settings (i.e. repetition time, TR, - and echo time, TE). The simulated FSE image $I_{FSE}$ is obtained as follows: An object descriptor class provides the physical properties of several tissue types $T$. In addition to information about the relative proton density $M_0$ and the relaxation parameters $T_1$ and $T_2$, it provides a logical mask indicating their disjunctive spatial distribution. For a given FA vector $\alpha$ and a predefined FSE sequence protocol including all relevant sequence parameters, the signal response $S$, i.e. the signal amplitude at each echo time in the echo train, is obtained for all tissue type via an EPG simulation: $S(T; \alpha) = (S_1(T; \alpha), \ldots, S_{ETL}(T; \alpha))$.

Using the $S$ of all tissue types and their spatial distribution, multi-contrast (MC) images are computed at the different echo times. After Fourier transform of the MC images, a single k-space is formed by combining k-space segments from different echo times according to the sampling table provided by the FSE sequence protocol. An inverse Fourier transform of the combined k-space then yields the FSE image $I_{FSE}$. Please note that only two dimensions are considered to reduce the computational complexity. The two dimensions consist of the phase encoding and partition encoding directions, respectively. The readout-dimension is ignored.

The PSF loss $L_{PSF}$ enforces reduced image blurring and is defined as (13)

$$L_{PSF}(\alpha) = \sum_{(T)issues}\sum_{i=1}^{N_y}\left(\frac{1}{N_y}c_T \cdot M \cdot PSF(T; \alpha)\right)_i \quad (4)$$

where $PSF(T; \alpha)$ is the point-spread-function of tissue $T$ for a given FAs $\alpha$, and $N_y$ is the matrix size in phase-encoding direction. The windowing function M is used to mask out the central voxel in $PSF$, because we only want to minimize its sidebands, as these lead to impurities from neighboring voxels in the imaging of the central voxel. The contribution of specific tissues to the loss can be controlled by the respective weighting factors $c_T \in [0,1]$. The point-spread functions $\boldsymbol{PSF}(T;\boldsymbol{\alpha})$ of each tissue type are obtained by the Fourier transform $\mathrm{F}(\cdot)$ of the respective modular transfer function $\boldsymbol{MTF}(T;\boldsymbol{\alpha})$:

$$\boldsymbol{PSF}(T;\boldsymbol{\alpha}) = \mathrm{F}(\boldsymbol{MTF}(T;\boldsymbol{\alpha})) \quad (5)$$

The $\boldsymbol{MTF}(T;\boldsymbol{\alpha})$ is obtained using $\boldsymbol{S}(T;\boldsymbol{\alpha})$ and the sampling table.

The regularization term $R(\boldsymbol{\alpha})$ ensures that there are no discontinuities between adjacent flip angles, and is defined by the $L_1$ distance of adjacent flip angles:

$$R(\boldsymbol{\alpha}) = \sum_{i=1}^{ETL} |\alpha_{1-i} - \alpha_i| \quad (6)$$

The FA vector $\boldsymbol{\alpha}$ is optimized via gradient descent using the ADAM optimizer (14) ($\beta_1 = 0.9$, $\beta_2 = 0.999$, learning rate: 0.001).

To consider the SAR constraint, we first note that all solutions $\boldsymbol{\alpha}$ satisfying the constraint live on a hypersphere of dimension $N = ETL$ and radius $R$, which is defined as

$$c_{SAR} = \frac{1}{ETL}\sum_{i=1}^{ETL}\left(\frac{\alpha_i}{180°}\right)^2 \Leftrightarrow c_{SAR} \cdot ETL \cdot 180°^2 := R = \sum_{i=1}^{ETL}(\alpha_i)^2 \quad (7)$$

Thus, starting with FAs $\boldsymbol{\alpha}$ on the hypersphere, $\boldsymbol{\alpha}$ can be projected onto the hypersphere after each gradient descent step simply by global rescaling, thereby ensuring that $\boldsymbol{\alpha}$ satisfies the SAR constraint in each optimization step. This optimization procedure is also known as projected gradient descent. The FAs $\boldsymbol{\alpha}$ were initialized by $\alpha_{1,\ldots,ETL} = \sqrt{R/ETL}$. The optimization framework was implemented in PyTorch (15), and is fully differentiable relative to the flip angles. Optimization was stopped when the loss of the following steps changed below a stopping criterion $c_{STOP} = 1e - 3$.

As 2D simulation phantom, we used the simulated brain dataset (SBD) (16)(17) to initialize the object descriptor with three different tissue types (white matter WM, gray matter GM, cerebrospinal fluid CSF. The physical properties used in this study were taken from reference (18).

### 2.2. Experiments

### Test case 1: T2w 3D FSE with Pulseq standard

The first imaging scenario is a 3D FSE with T2 weighting. A 3D FSE sequence was implemented within the open source sequence development platform PyPulseq (19) (20) (interpreter version 1.3.1post1). In the spirit of reproducible research, all optimized sequences are accessible at our Github repository: https://github.com/MRsources/hyperSpace. The sequence parameters were as follows: repetition time TR=3200ms, echo spacing ESP=3.8ms, effective echo time TE_eff=380.0 ms, FOV$_x$ x FOV$_y$ x FOV$_z$=300 x 200 x 50 mm³, resolution= 1.5 x 1 x 1 mm³, echo train length ETL=200, rel.SAR c$_{SAR}$=9.0%, bandwidth=234Hz/pixel.

Linear view ordering was used in the phase-encoding direction, and partitions were encoded linearly with one shot per partition (i.e., one TR per partition). The total acquisition time was 160.0s.

Two optimization schemes were considered: VFA optimization to minimize image blurring and to maximize SNR, i.e. $w = 1$ and $w = 0$, respectively in Eq. (1). All tissues were equally weighted in the PSF loss (i.e. $c_T = 1$ in Eq. (4) for T=WM, GM and CSF). The spin echo target image $I_{TRG}$ was assigned an echo-time of TE = 45 ms and a repetition time of TR = 6000 ms.

In vivo measurements were performed on two healthy volunteers using a 7T whole-body MR system (Magnetom Terra.X, Siemens Healthineers, Erlangen, Germany) with a commercial 8Tx/32Rx RF head coil. All MRI scans were under approval of the local ethics board, and were performed after written informed consent was obtained.

### Quantitative PSF comparison

For quantitative estimation of image blurring, unencoded signals $S$ were acquired for each investigated VFA scheme $\alpha_{VFA}$ (PSF- and SNR-optimized as well as standard). To this end, the partition- and phase-encoding gradients were turned off, and the ADC data were acquired after a dummy excitation with one echo train. The echo tops of the readouts (coil averaged) were assigned as $S$. To ensure that most of the contributions to the unencoded signal originated from white and gray matter, we used a dark-fluid inversion recovery pulse with the inversion time TI adjusted to suppress the CSF signal ($TI = 1800$ ms). The obtained $S$ for each VFA scheme were then used to calculate the corresponding, tissue-averaged PSF. As reference PSF, $PSF_{REF}$, we used the PSF of a constant target signal, i.e. $S_{REF} = 1 \ \forall \ i = 1, \dots, ETL$. The $PSF$ of the respective VFA schemes $\alpha_{VFA}$ was then subtracted from $PSF_{REF}$, and the residual error $ERR_{\alpha_{VFA}}$ was defined as:

$$ERR_{\alpha_{VFA}} = \frac{\sum_{i=1}^{N_y} |(PSF_{REF} - PSF_{\alpha_{VFA}})|_i}{\sum_{i=1}^{N_y} |(PSF_{REF})|_i} \quad (8)$$

where $N_y$ denotes the matrix size in phase-encoding direction.

### Quantitative SNR comparison

To estimate the SNR for each investigated VFA scheme $\alpha_{VFA}$, Monte Carlo simulations (also known as pseudo-replica method) were performed (21). To this end, the respective k-space data were pre-whitened (i.e. transformed into uncorrelated coils with equal noise level) using a noise-only scan. The image reconstruction process, which in this case consists only of a Fourier-transform, is then repeated $N_{REP} = 100$ times, where for each repetition, with the pre-whitened k-space data being independently superimposed with Gaussian noise. For each voxel $k$, the SNR is then defined as the mean $\mu^{(k)}$ over the standard deviation $\sigma^{(k)}$ over the noise-perturbed pseudo-replicas:

$$SNR_{\alpha_{VFA}}^{(k)} = \frac{\mu_{\alpha_{VFA}}^{(k)}}{\sigma_{\alpha_{VFA}}^{(k)}} \quad (9)$$

### Test case 2: T2w 3D FSE with vendor sequence

To evaluate the applicability of the PSF- and SNR-optimized FA schemes in clinical routine, they were tested beyond the Pulseq standard in a vendor sequence. The vendor sequence was modified so that an external list of VFAs can be loaded. The vendor sequence was assigned the same imaging parameters for contrast setting as for the Pulseq sequence in test case 1 (i.e. TR, ESP, $TE_{eff}$) and the same ETL and rSAR. The FOV and resolution were set to $FOV_x \times FOV_y \times FOV_z = 230 \times 230 \times 180 \text{ mm}^3$, resolution $= 0.4 \times 0.4 \times 0.5$ (interpolated). In addition, parallel imaging was performed using CAIPIRINHA (22) with a total of 6-fold acceleration (PE:3, 3D:2, shift:1). The total acquisition time was 279.0 seconds. Furthermore, a significant difference to the Pulseq sequence is that this vendor sequence utilizes optimized pTx pulses to mitigate signal variations throughout the FOV due to B1 inhomogeneities (23)(24).

In vivo measurements were performed on two healthy volunteers using the same experimental setup described in test case 1.

### Test case 3: 3D FLAIR with vendor sequence

In the third test case, a 3D vendor FSE with T2w-FLAIR weighting was considered. The imaging parameters were as follows: TR=8000ms, ESP=4.04ms, TE_eff=300.0ms, TI= 2250ms, $FOV_x \times FOV_y \times FOV_z$=230 x 230 x 180 mm³, resolution= 0.4 x 0.4 x 0.5 mm³ (interpolated), ETL=220, rel. SAR $c_{SAR}$=6.0%, bandwidth=651Hz/pixel, linear reordering scheme, acceleration with CAIPIRINHA (total factor 6,PE: 3, 3D: 2, shift: 1). The total acquisition time is 353.0 seconds.

During optimization, the CSF compartment in the spin-echo target image $I_{TRG}$ was suppressed in this imaging scenario to account for fluid suppression. Correspondingly, an additional inversion pulse was included in the EPG simulation with a time delay of TI to the 90° excitation pulse.

VFA optimization to minimize image blurring and to maximize SNR, i.e. w=1 and w=0 in Eq. (1), respectively, was considered. All tissues were equally weighted in the PSF loss (i.e., $c_T$=1 in equation (4) for T=WM and GM). The spin echo target image $I_{TRG}$ was assigned an echo time of TE = 45 ms and a repetition time of TR = 6000 ms.

In vivo measurements were performed on one healthy volunteer using the same experimental setup described in test case 2.

3. Results

### Test case 1: VFA schemes

Figure 2 shows the VFA schemes and the corresponding signal response simulations and PSFs for the standard (Figure 2. A) and the PSF and SNR optimizations (Figure 2. B and C, respectively). The PSF optimization results in a VFA scheme that is similar to the standard scheme, however, with significant differences. The flip angles in the first three quarters of the flip angle train are reduced, and elevated for the last quarter. Consequently, the pseudo-steady state signal responses of WM and GM are elongated compared to the signal response of the standard VFA scheme. This results in corresponding PSFs with greatly reduced sidebands compared to the corresponding PSFs of the standard VFA scheme. While the PSFs of CSF, WM, and GM deviate in the standard, they are equalized by the PSF optimization. Note that the resulting signal gap from CSF to WM and GM at the effective echo time is reduced. This allows for images with more regularized CSF signal relative to WM and GM compared to the standard, while still preserving contrast (see next section, Figure 3).

On the other hand, the SNR optimization yields a VFA scheme that is comparable to the PSF-optimized VFA in the first two quarters. However, it shows a characteristic peak at the effective echo time. This results in elevated signal responses at the echoes assigned to the k-space center, which are accompanied by elevated sidebands in the corresponding PSFs that exceed those of the standard, which results in enhanced image blurring (see next section, Figure 3).

### Test case 1: Pulseq 3D FSE Imaging

Figure 3 shows an exemplary transversal partition of the images obtained with the Pulseq sequences using the standard and the optimized VFA schemes shown in Figure 2 A for one subject. Compared to the standard, the PSF-optimized VFA scheme results in greatly reduced image blurring in the phase-encoding direction (anterior > posterior), while providing only a slightly altered contrast. In particular, WM/GM structures are sharper and WM/GM edges are better resolved. In addition, blood vessels are more pronounced, as shown in the zoomed region (see Figure 2, insets). This is partly due to a more regularized CSF signal relative to WM and GM due to the PSF optimization, but can also be explained by less GM/WM blur into low intense vessel regions. On the other hand, the SNR-optimized VFA scheme provides increased signal values, however, at the cost of increased visual image blurring while providing same contrast setting. It is worth noting that the relative SAR is equal for all VFA schemes.

Figure 4 shows a 3D view of this test case on subject 2. It is worth noting that the PSF-optimized variant also shows reduced image blurring in the axial and sagittal planes. Note this sequence did not incorporate optimized pTx pulses to mitigate B1 inhomogeneities. Thus, signal variations throughout the brain are present. However, the compatibility of the optimized VFA schemes with optimized pTx pulses is demonstrated using vendor sequences in test case 2 and 3.

### Quantitative PSF comparison

After visual inspection, a quantitative comparison of PSF-induced image blurring is shown in Figure 5. The sequence from test case 1 was applied without partition and phase encoding gradients on subject 1 using the standard VFA and both PSF- and SNR-optimized VFAs. The resulting unencoded signals are shown in Figure 5 A. It is worth noting that the experimental, in vivo signal responses are consistent with the simulated signal response of WM and GM in all cases (see Figure 5 A and signal responses in Figure 1). This is due to the minimization of CSF signal contributions using an inversion pulse with an inversion time that was adjusted to suppress the CSF signal.

The corresponding PSFs are depicted in Figure 5 B. The PSF optimization results in a PSF with strongly reduced sidebands compared to the PSF of the standard VFA. On the other hand, the SNR optimization shows elevated sidebands. In Figure 5 C, the error to the sinc-like reference

PSF are depicted. Using Eq. 8, the error was quantified (Figure 5 C). While the error of the standard amounts to 1.27, the PSF optimization yields a minimum error reduced by 51.2 % (0.62). On the other hand, the SNR optimization reveals a maximum error (1.55) which increases the standard by 22.0 %

Quantitative SNR comparison

In addition to the PSF quantification, the SNR was analyzed quantitatively using Monte Carlo simulations to obtain SNR maps for both subjects (see Figure 6). Within a WM/GM ROI, the PSF optimization yields an average SNR which is decreased by 37.4% relative to the standard for subject 1 (25.9±5.5 vs. 41.4±11.5, respectively). On the other hand, the SNR optimization yields an average SNR that is increased by 96.1% relative to the standard (81.2±18.4 vs. 41.4±11.5, respectively). Note that similar observations were obtained for subject 2: The PSF optimization shows decreased average SNR (reduced by 44.6% relative to standard (22.1±4.4 vs. 39.9±11.0, respectively), while the SNR optimization yields an average SNR elevated by 83.0% relative to the standard (73.0±15.6 vs 39.9±11.0, respectively).

Test case 2: T2w 3D FSE with vendor sequence

The VFA schemes from test case 1 were used in the corresponding vendor sequence with ptX extension. The results for subject 3 and subject 4 are shown in Figures 7 and 8, respectively. In general, the observations from the images obtained via the Pulseq standard in test case 1 are being validated for both subjects: In both examinations, the PSF-optimized VFA scheme shows strongly reduced image blurring relative to the standard, in particular for WM/GM structures. The blood vessels in axial view (Figure 7 insets), and small structures in the cerebellum in sagittal view (Figure 8 insets) are better resolved. Note that the blurring of the cochlea is strongly reduced, too, which is particularly helpful for segmentation tasks (Figure 8 insets). The contrast fidelity in comparison to the standard is well preserved. It is worth noting that this imaging scenario employs an echo train duration of 760.0ms, and the PSF optimization still yields highly resolved WM and GM structures with a reasonable SNR, which is worth emphasizing as WM and GM typically have T2 relaxation times of 40-50ms and 50-60ms at 7T, respectively.

The SNR optimization, on the other hand, strongly amplifies signal magnitudes, while maintaining contrast settings and relative SAR. However, in accordance to the visual inspections via the Pulseq sequence in test case 1 and quantitative analysis of image blurring, the SNR-enhanced images show elevated image blurring in this case, too. It should be noted that all VFA schemes have equal relative SAR.

The used pTx RF pulses yield homogenous signal throughout the brain, and signal variations caused by the B1 inhomogeneities, as present in the Pulseq acquisition (Figure 4), are strongly mitigated by the vendor acquisition. As we observe the same effects of VFA, as with the 1Tx Pulseq version, we can conclude that our VFAs are compatible with pTx pulses.

Test case 3: 3D FLAIR with vendor sequence

In the third test case, flip angles were optimized for a T2-FLAIR sequence. The resulting images obtained using a vendor sequence with pTx RF pulses are shown in Figure 9 along with the corresponding VFA schemes (Figure 9, right had side). This experiment shows that the T2-FLAIR sequence also benefits from the optimizations: The PSF-optimized images display highly resolved WM and GM structures and reduced image blurring while providing high contrast fidelity and reasonable SNR. Note that the total echo train length is 220 in this case, resulting into an echo train duration of 888.8 ms. Highly amplified signal magnitudes are provided by the SNR-optimized VFA scheme. Note that the k-space center is assigned the 75th echo, and the SNR-optimized VFA scheme reveals elevated flip angles around this echo number. It should be noted that all VFA schemes have equal relative SAR.

## 4. Discussion

The proposed framework builds on the initial results of conference contributions (13)(25) and introduces a flexible and robust method for the automatic optimization of VFA schemes in 3D FSE sequences, addressing key challenges related to SNR and PSF penalties in case of very long echo trains at 7T. By adjusting the weighting of cost function components, VFA schemes were derived automatically that favor sharpness or SNR at equal relative SAR. PSF-flavored schemes provided better-resolved white and gray matter structures in both T2w and FLAIR sequences in trade-off for SNR, while SNR-flavored schemes provided increased signal levels in trade-off for sharpness.

Automatic parameter optimization for FSE sequences has been investigated in previous works. Dang et al. optimized VFA schemes end-to-end within the MR–zero framework (9)(10) for single-shot 2D FSE sequences with PDw and T2w at 3T. T2 blurring was significantly reduced and the optimized VFA schemes outperformed schemes proposed by Zhao et al., where FA trains were optimized by performing signal simulation to produce a desired signal response (11), and TRAPS (3) However, while this work showed the benefit of end-to-end VFA optimization over standard approaches, the considered sequences were limited to 2D acquisitions at 3T with echo train lengths of 64. This work makes principal contributions by extending towards 3D acquisitions at 7T with very long echo trains, for which a dedicated optimization framework is proposed. It should be noted that Dang et al. used a convolutional neural network as a deblurring network (26), and the magnitude FSE image was assigned as input. It was trained jointly to the VFA optimization. In this work, however, we did not employ a deblurring network, as the aspects of network generalization and the substantial demand for training data and time remain to be elucidated. Instead, the objective of reducing image blurrig was achieved via an explicit, dedicated loss term, which is minimized in the VFA optimization (13). While this approach eliminates the computational burden associated with CNN training for 3D data, incorporating a deblurring network into the proposed framework is possible as Dang et al. proposed. Optimizing VFAs with explicit PSF loss term and a deblurring network jointly promises further enhancements for the visibility of small structures.

De Buck et al. employed a comprehensive DANTE-SPACE simulation framework for the optimization of T2-weighted DANTE-SPACE at both 7T and 3T (8). The simulation-based optimization of the DANTE parameters facilitated improved T2w DANTE-SPACE contrast at 7T. However, the sequence optimization was limited to DANTE parameters, and optimizing both the VFA scheme and DANTE parameters jointly remains to be elucidated, and could lead to further improvements.

This work addresses a critical gap in prior research by optimizing the PSF across multiple tissue types. Previous VFA schemes often focused on single tissue optimization where single relaxation parameters are considered (3)(4)(7)(11)(27)(28). In contrast, by incorporating the optimization of the PSF for multiple tissue types simultaneously, more balanced signal levels of WM/GM and CSF were achieved in the T2w sequences. This significantly reduces CSF contamination, a common issue that obscures structural details, particularly at tissue interfaces. Indeed, in the PSF-optimized T2w images, enhanced visibility of blood vessels was achieved for both the Pulseq- and the vendor sequence. The improved visibility of small structures in the T2w images by PSF-optimized VFAs may facilitate and improve common segmentation tasks where T2w 3D FSE sequences are used (29)(30).

While the proposed framework provides a flexible tool for optimizing both PSF and SNR, the balance between these two objectives should be carefully adjusted based on clinical priorities. A higher weighting on PSF optimization improves spatial resolution but may come at the cost of reduced SNR, potentially affecting the visibility of fine details in certain tissues. Conversely, prioritizing SNR optimization can be beneficial where the base-line SNR is low, e.g. in thorax imaging, as it enhances signal intensity but may introduce blurring artifacts. Therefore, determining the optimal trade-off between PSF and SNR is not purely a technical decision but requires input from a radiologist. It is essential to ensure that the resulting images align with the clinical objectives of the scan, whether that be sharper delineation of tissue boundaries or maximized signal intensity for better contrast.

One of the strengths of this framework lies in its versatility, allowing the optimization process to be tailored to specific tissues or anatomical regions. In applications where high-resolution depiction of small structures is required, like inner ear or cochlea imaging (31), the optimization can be weighted to prioritize the PSF for that particular region. This capability is especially useful in cases where certain tissues demand higher resolution than others due to their small size or critical diagnostic importance. By fine-tuning the optimization for specific tissues or regions of interest, clinicians can extract the maximum diagnostic value from each scan, without compromising the overall image quality for other regions.

Another advantage is the scalability to include additional tissue types or even pathological structures. For example, multiple sclerosis (MS) lesions, which can be difficult to visualize due to their small size and diffuse nature, can be included in the optimization process through the object descriptor. This extension would allow for tailored flip angle schemes that prioritize the visualization of pathological areas. It is relatively straightforward to adapt it for imaging protocols targeting other pathologies, such as tumors or ischemic regions, where clear differentiation from surrounding tissue is critical.

### Limits

The model employed in this work was deliberately simplified to focus on the core objective of flip angle optimization. More complex effects such as magnetization transfer (MT) were not incorporated. While this allowed for faster computation, we acknowledge that MT effects, particularly in multi-slice acquisitions, can have a substantial impact on image contrast and are also dependent on the VFA scheme. These effects should not be overlooked, as MT can alter the effective relaxation times and contribute to variations in contrast, especially in tissues with high macromolecular content. This can be achieved by extending the EPG with MT effects (32).

Scanner imperfections such as B0 and B1 inhomogeneities can degrade the performance of VFA-optimized sequences, particularly at higher field strengths where these inhomogeneities are more pronounced. To address these limitations, we plan to incorporate the MR-zero framework (10) into the optimization pipeline.

To achieve practical computational performance, a simplified optimization model was used that operates on a "tissue-wise" rather than a pixel-wise basis. By focusing on representative tissue properties rather than simulating every individual pixel, the computational cost was significantly reduced while still maintaining a high degree of accuracy in predicting optimal flip angles. Additionally, by employing 2D simulations instead of full 3D modeling, we further streamline the process, enabling rapid computation of VFA schemes.

### Future works

The incorporation of more sophisticated loss terms such as the CNR could provide further improvements in contrast optimization. The framework could more effectively balance image contrast and noise characteristics, particularly in sequences that require high contrast differentiation between tissues (e.g. gray and white matter in neuroimaging or cartilage and bone in musculoskeletal imaging). Furthermore, incorporating tissue-wise relative contrast constraints (e.g. lesion-to-white/gray matter) could prove beneficial for imaging crucial diagnostic structures.

In knee imaging, optimizing VFAs could enhance the visualization of cartilage and meniscal structures. Cochlea imaging is another subject, where the high spatial resolution and tissue contrast provided by optimized FSE sequences could contribute to better characterization of inner ear structures. It is worth noting that slab-selective FSE (5)(33) is typically used for inner ear imaging (34). In this work, whole-brain imaging was considered. However, future works will investigate combining the flexible VFA scheme optimization with slab-selective excitation, including dedicated considerations about transmit field inhomogeneities. Furthermore, integration of kT spokes optimization (35) for further refinement of spatial selectivity and resolution could prove beneficial. In this regard, hippocampal imaging is another promising subject where the PSF- or SNR-optimized VFA schemes in T2w sequences could provide better resolved structures for manual segmentation (36).

The fast and flexible adjustment of the VFA flavor in real time at the scanner, depending on the task and the tissues of interest, is another important aspect. In this work, one iteration step costs approx. 20.3 seconds, and the total iteration number were 252 for the T2w test case and 191 for the FLAIR test case. As demonstrated in this study, the VFAs can be tailored to favor either the PSF or the SNR, with each approach offering distinct advantages. A PSF-like VFA may prove advantageous for the identification of small structures in the knee, hippocampus, or the identification of MS lesions. Conversely, a SNR-like VFA may be advantageous in imaging regions where low baseline SNR is anticipated, such as the thorax. As shown in this work, the desired flavor can be achieved by weighting the PSF- and SNR-loss terms in the loss function.

5. Conclusion

This study demonstrates the potential of end-to-end learning frameworks to optimize VFA schemes in very long echo trains for 3D FSE acquisition at 7T in terms of PSF and SNR. By adjusting the weighting of cost function components, VFA schemes were obtained which favor sharpness (PSF), or increased signal levels (SNR). It paves the way for the fast and flexible adjustment of the trade-off between PSF and SNR for 3D FSE imaging, depending on the specific task. PSF-like VFAs are more beneficial for the optimized imaging of small details, while SNR-like VFAs are more beneficial where low base-line SNR deteriorates image quality.

### Conflict of interest statement

Jürgen Herrler and Patrick Liebig are employees of Siemens Healthineers (Erlangen, Germany).

### Data availability statement

In the spirit of reproducible research, all optimized sequences are accessible at our Github repository: https://github.com/MRsources/hyperSpace

### References

1) Hennig J, Nauerth A, Friedburg H. RARE imaging: a fast imaging method for clinical MR. Magn Reson Med. 1986 Dec;3(6):823-33.
2) Bottomley PA, Redington RW, Edelstein WA, Schenck JF. Estimating radiofrequency power deposition in body NMR imaging. Magn Reson Med. 1985 Aug;2(4):336-49.
3) Hennig J, Weigel M, Scheffler K. Multiecho sequences with variable refocusing flip angles: optimization of signal behavior using smooth transitions between pseudo steady states (TRAPS). Magn Reson Med. 2003 Mar;49(3):527-35.
4) Busse RF, Hariharan H, Vu A, Brittain JH. Fast spin echo sequences with very long echo trains: design of variable refocusing flip angle schedules and generation of clinical T2 contrast. Magn Reson Med. 2006 May;55(5):1030-7.
5) Mugler JP. Optimized three-dimensional fast-spin-echo MRI.J Magn Reson Imaging. 2014;39:745-767.
6) Weigel M, Hennig J. Contrast behavior and relaxation effects of conventional and hyperecho-turbo spin echo sequences at 1.5 and 3 T. Magn Reson Med. 2006 Apr;55(4):826-35. Erratum in: Magn Reson Med. 2009 Jun;61(6):1540
7) Keerthivasan, M.B., Galons, J.-P., Johnson, K., Umapathy, L., Martin, D.R., Bilgin, A. and Altbach, M.I. (2022), Abdominal T2-Weighted Imaging and T2 Mapping Using a Variable Flip Angle Radial Turbo Spin-Echo Technique. J Magn Reson Imaging, 55: 289-300.
8) de Buck MHS, Hess AT, Jezzard P. Simulation-based optimization and experimental comparison of intracranial T2-weighted DANTE-SPACE vessel wall imaging at 3T and 7T. Magn Reson Med. 2024; 92: 2112-2126.
9) Dang HN, Endres J, Weinmüller S, et al. MR-zero meets RARE MRI: Joint optimization of refocusing flip angles and neural networks to minimize T2-induced blurring in spin echo sequences. Magn Reson Med. 2023; 90(4): 1345-1362.
10) Loktyushin A, Herz K, Dang N. MRzero - Automated discovery of MRI sequences using supervised learning. *Magn Reson Med*. 2021; 86: 709–724.
11) Zhao L, Chang C-D, Alsop DC. Controlling T2 blurring in 3DRARE arterial spin labeling acquisition through optimal combi-nation of variable Flip angles and k-space filtering. Magn ResonMed. 2018;80:1391-1401
12) Hennig J. Multiecho imaging sequences with low refocusing flip angles. J Magn Reson 1988; 78: 397–407

13) Dawood P, Blaimer M, Weinmüller S, Jakob PM, Knoll F, Zaiss M. End-to-end optimization of 3D-TSE with very long echo trains. Annual Meeting of the ESMRMB, Barcelona, ES; Abstract number 0113.
14) Kingma DP, Ba J. Adam: a method for stochastic optimiza-tion. Proceedings of 3rd International Conference on LearningRepresentations. ICLR; 2015.
15) Paszke A, Gross S, Massa F, et al. PyTorch: an imperative style, high-performance deep learning library. Proceedings of the 33rdConference on Neural Information Processing Systems; 2019
16) Collins DL, Zijdenbos AP, Kollokian V, Sled JG, Kabani NJ, Holmes CJ, Evans AC. Design and Construction of a Realistic Digital Brain Phantom. IEEE Transactions on Medical Imaging. 1998;17(3):463—468.
17) Kwan RKS, Evans AC, Pike GB. MRI simulation-based evaluation of image-processing and classification methods. IEEE Transactions on Medical Imaging. 1999;18(11):1085-97.
18) Zhu J, Klarhöfer M, Santini F, Scheffler K, and Bieri O. Relaxation Measurements in Brain tissue at field strengths between 0.35T and 9.4T. Proceedings of the 22th Annual Meeting of ISMRM; 2014 Abstract Program Number 3208.
19) Layton, Kelvin J., et al. "Pulseq: a rapid and hardware-independent pulse sequence prototyping framework." Magnetic resonance in medicine 77.4 (2017): 1544-1552.
20) Ravi, Keerthi, Sairam Geethanath, and John Vaughan. "PyPulseq: A Python Package for MRI Pulse Sequence Design." Journal of Open Source Software 4.42 (2019): 1725.
21) Robson, P.M., Grant, A.K., Madhuranthakam, A.J., Lattanzi, R., Sodickson, D.K. and McKenzie, C.A. (2008), Comprehensive quantification of signal-to-noise ratio and g-factor for image-based and k-space-based parallel imaging reconstructions. Magn. Reson. Med., 60: 895-907.
22) Breuer, F.A., Blaimer, M., Heidemann, R.M., Mueller, M.F., Griswold, M.A. and Jakob, P.M. (2005), Controlled aliasing in parallel imaging results in higher acceleration (CAIPIRINHA) for multi-slice imaging. Magn. Reson. Med., 53: 684-691.
23) Herrler J, Liebig P, Gumbrecht R, et al. Fast online-customized (FOCUS) parallel transmission pulses: A combination of universal pulses and individual optimization. Magn Reson Med. 2021; 85: 3140–3153.
24) Herrler J, Majewski K, Liebig PA, Benner T, Ferguson GW, Gumbrecht R, Insua IG, and Heidemann RM. Scalable fast online-customized (FOCUS) pTx pulses for 3D TSE sequences at 7T. Proceedings of the 32th Annual Meeting of ISMRM; 2024 Abstract Program Number 4097.
25) Blaimer M, Breuer FA, Weber D, Malik SJ, Herold V. Simulation frame work for predicting flip angles in echo-train sequences. Annual Meeting of the ESMRMB, Rotterdam, NL; Magn Reson Mater Phy 2019; 32(Suppl 1) 21.
26) Xu L, Ren JSJ, Liu C, Jia J. Deep convolutional neural networkfor image deconvolution. Proceedings of the 27th InternationalDANG et al. 1361Conference on Neural Information Processing Systems – Volume1 (NIPS'14). MIT Press; 2014:1790-1798.
27) Busse RF, Brau ACS, Vu A, et al. Effects of refocusing flip angle modulation and view ordering in 3D fast spin echo. Magn ResonMed. 2008;60:640-649.
28) Rahbek S, Schakel T, Mahmood F, Madsen KH, PhilippensMEP, Hanson LG. Optimized flip angle schemes for the splitacquisition of fast spin-echo signals (SPLICE) sequence andapplication to diffusion-weighted imaging. Magn Reson Med.2023;89:1469-1480.
29) Helms, G., Kallenberg, K. and Dechent, P. (2006), Contrast-driven approach to intracranial segmentation using a combination of T2- and T1-weighted 3D MRI data sets. J. Magn. Reson. Imaging, 24: 790-795.
30) S. Farcito et al., "Accurate anatomical head segmentations: a data set for biomedical simulations," 2019 41st Annual International Conference of the IEEE Engineering in Medicine and Biology Society (EMBC), Berlin, Germany, 2019, pp. 6118-6123
31) Benson JC, Carlson ML, Lane JI. MRI of the Internal Auditory Canal, Labyrinth, and Middle Ear: How We Do It. Radiology 2020 297:2, 252-265

32) Malik SJ, Teixeira RPAG, Hajnal JV. Extended phase graph formalism for systems with magnetization transfer and exchange. Magn Reson Med. 2018 Aug;80(2):767-779.
33) Park, J., Mugler, J.P., III and Hughes, T. (2009), Reduction of B1 sensitivity in selective single-slab 3d turbo spin echo imaging with very long echo trains. Magn. Reson. Med., 62: 1060-1066.
34) Floc'h JL, Tan W, Telang RS, Vlajkovic SM, Nuttall A, Rooney WD, Pontré B, Thorne PR. Markers of cochlear inflammation using MRI. J Magn Reson Imaging. 2014 Jan;39(1):150-61.
35) Lowen D, Pracht ED, Gras V, et al. Design of calibration-free RF pulsesfor T2 -weighted single-slab 3D turbo-spin-echosequences at 7T utilizing parallel transmission. Magn Reson Med. 2024;92:2037-2050
36) Berron D, Vieweg P, Hochkeppler A, et al. A protocol for manual segmentation of medial temporal lobe subregions in 7 tesla MRI. Neuroimage Clin. 2017;15:466-482.

## Figures

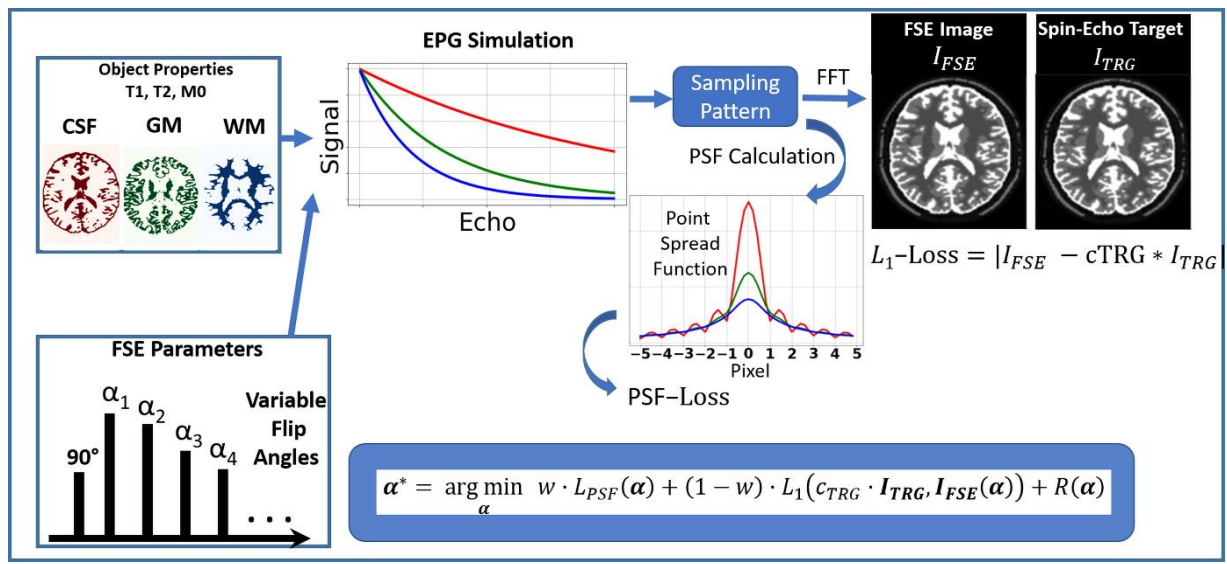

Figure 1

The optimization framework comprises an object descriptor indicating the physical properties (T1, T2, M0) and the disjunct spatial distribution of the tissue types of interest. Given a FSE sequence protocol and variable flip angles, an EPG simulation provides the signal response for each tissue type. Using the view ordering (i.e. sampling table), a k-space is formed from which an image is obtained by a Fourier transformation. Furthermore, the signal responses provide the point-spread-function (PSF) of all tissue types using the view ordering. For optimization of variable flip angles, a loss function is defined that includes the L1-error of the simulated FSE image relative to a target image (spin-echo image with predefined repetition- and echo-time), which serves as a SNR loss that provides contrast fidelity. Furthermore, a PSF-error is defined. The weighting factor controls the contributions of SNR-and PSF-loss. An additional regularization term R is used to enforce smooth signal courses.

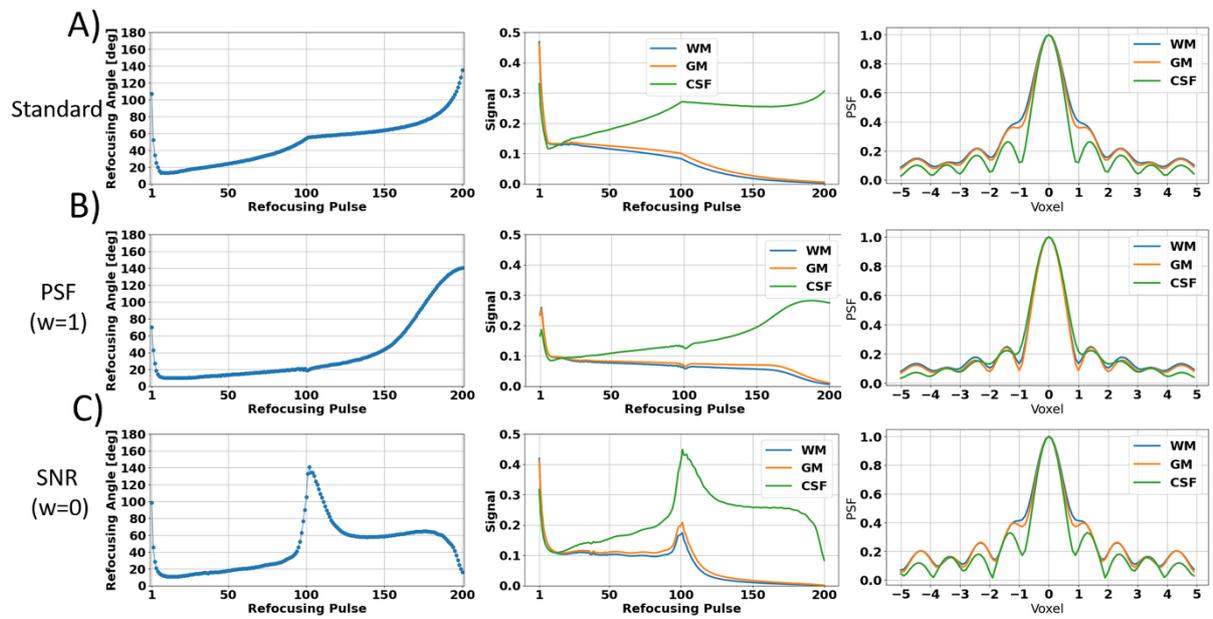

Figure 2

(A) The VFA scheme (left), simulated signal responses (center) and PSFs (right) of the standard approach. The signal decays exponentially, keeps a pseudo steady-state until the echoes assigned to k-space center are reached, and then decays exponentially. (B) depicts the PSF-optimized case, where the pseudo steady-state period is elongated in comparison to (A), resulting in PSFs of white and gray matter showing strongly reduced side bands minimizing image blurring. (C) depicts the SNR-optimized case. A characteristic peak in the VFA scheme and signal response at the echoes assigned to k-space center originates, leading to enhanced SNR and increased PSF side bands. Note that the relative SAR is kept constant in all cases.

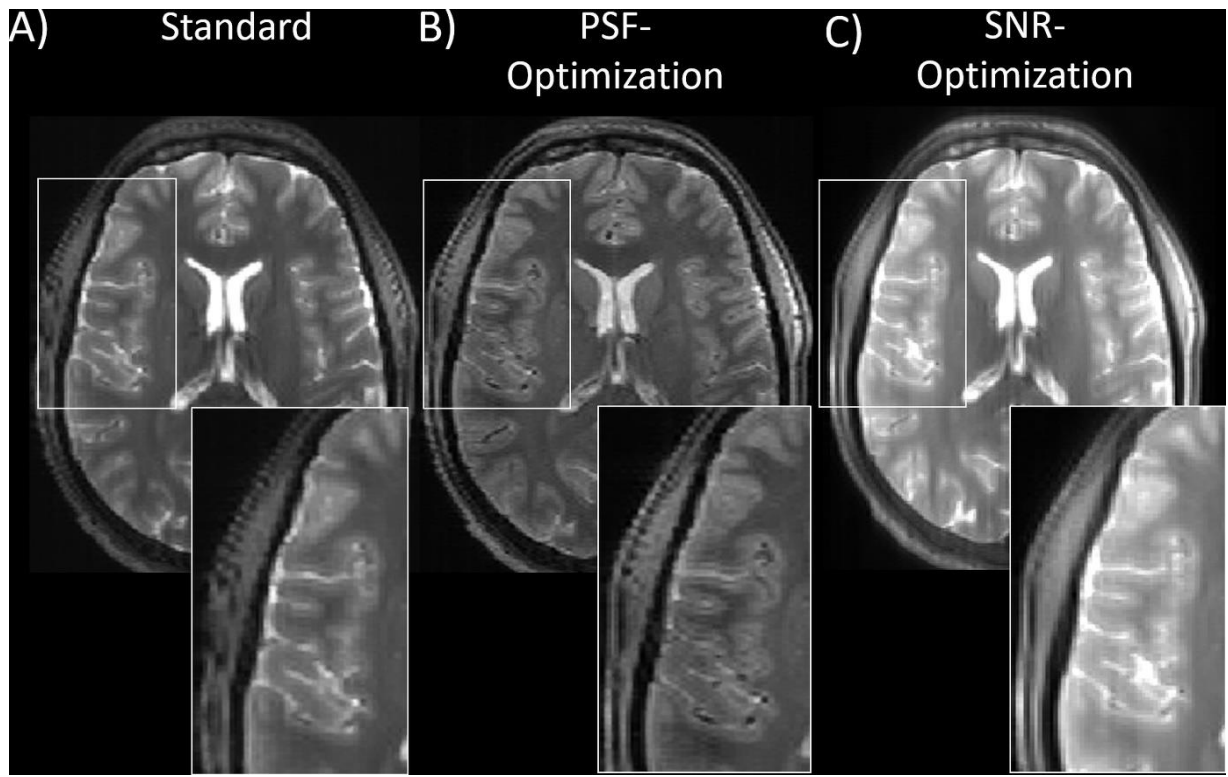

Figure 3

Images of subject 1 obtained via the Pulseq 3D FSE sequence with T2w for standard, PSF- and SNR-optimized VFA schemes (A-C, respectively). In comparison to the standard, the PSF-optimized scheme yields images with better resolved white- and gray matter structures while providing contrast fidelity. On the other hand, the SNR-optimized scheme yields increased signal magnitudes which come along with increased image blurring. Note that this sequence does not incorporate parallel transmit and all cases are assigned the same relative SAR.

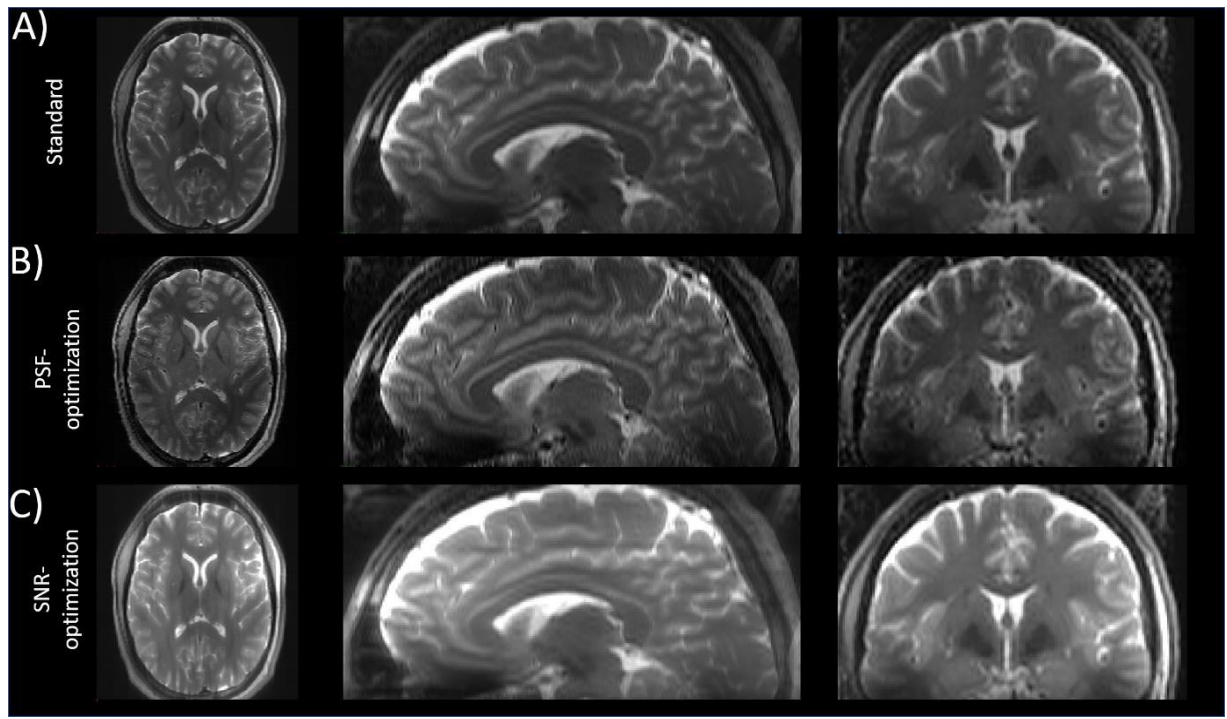

Figure 4

3D view of images of subject 2 obtained via the Pulseq 3D FSE sequence with T2w for standard, PSF- and SNR-optimized VFA schemes (A-C, respectively). Better resolved white- and gray matter structures in the PSF-optimization are visible in axial, transversal and sagittal planes, while the SNR optimization shows increased signal magnitudes and enhances image blurring. Note that in this sequence, no parallel transmit has been used, thus, signal variations throughout the FOV are visible in all cases. It should be also noted that all VFA schemes have equal relative SAR.

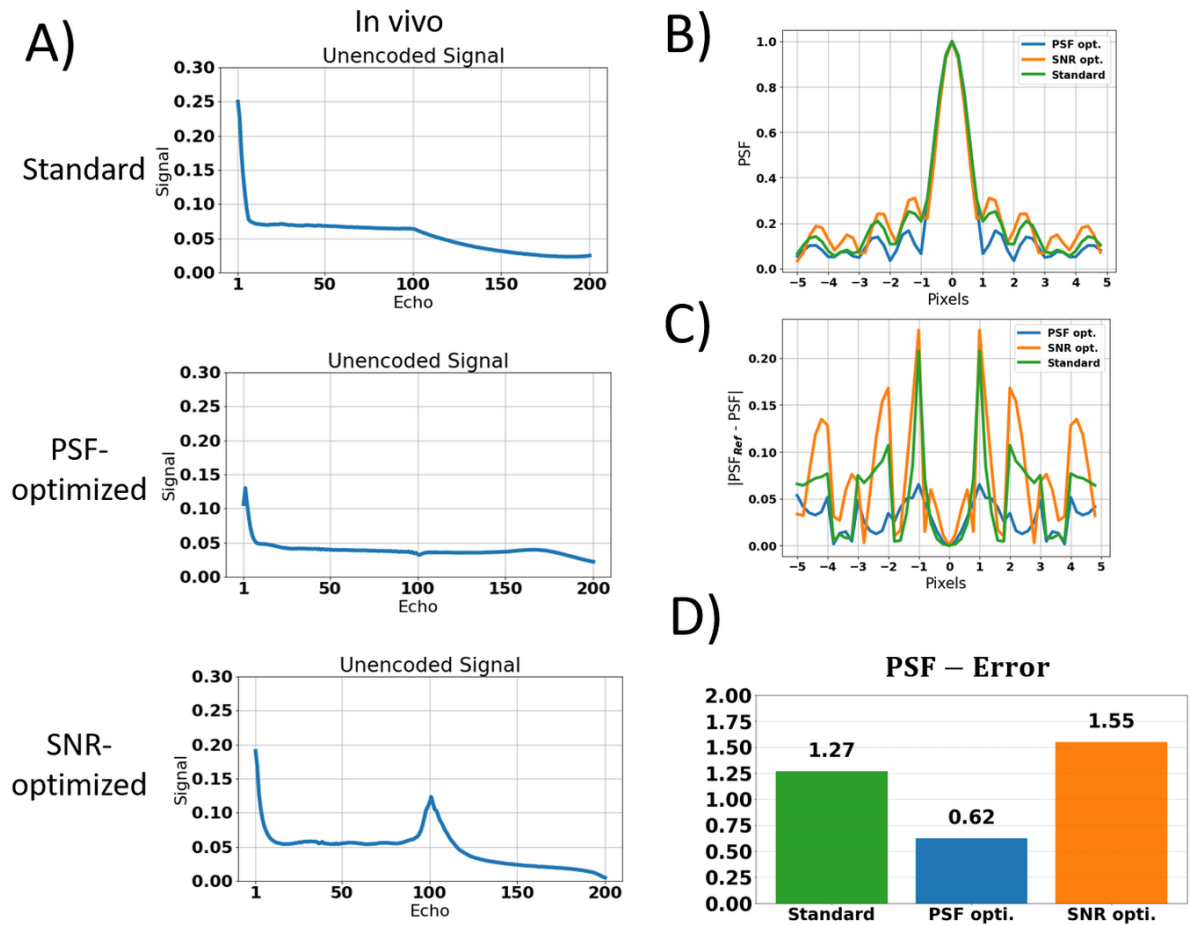

Figure 5

(A) The unencoded signal responses for standard, PSF- and SNR-optimized VFA schemes for the T2w 3D FSE sequence. Note that the CSF contribution has been suppressed using an inversion pulse. The resulting invivo signal responses are in good agreement with those of white- and gray matter from the simulations (Figure 2). (B) The PSFs computed form the unencoded invivo signals (A). (C) The differences of the invivo PSFs from a reference, sinc-like PSF, which corresponds to a uniform signal response over the echo train. (D) The quantitative error (i.e. the normalized L1-error to the sinc-like reference) of standard, PSF- and SNR-optimized VFA schemes. Note that the PSF-optimization resulted in sidebands reduced by approx. 50% relative to the standard VFA.

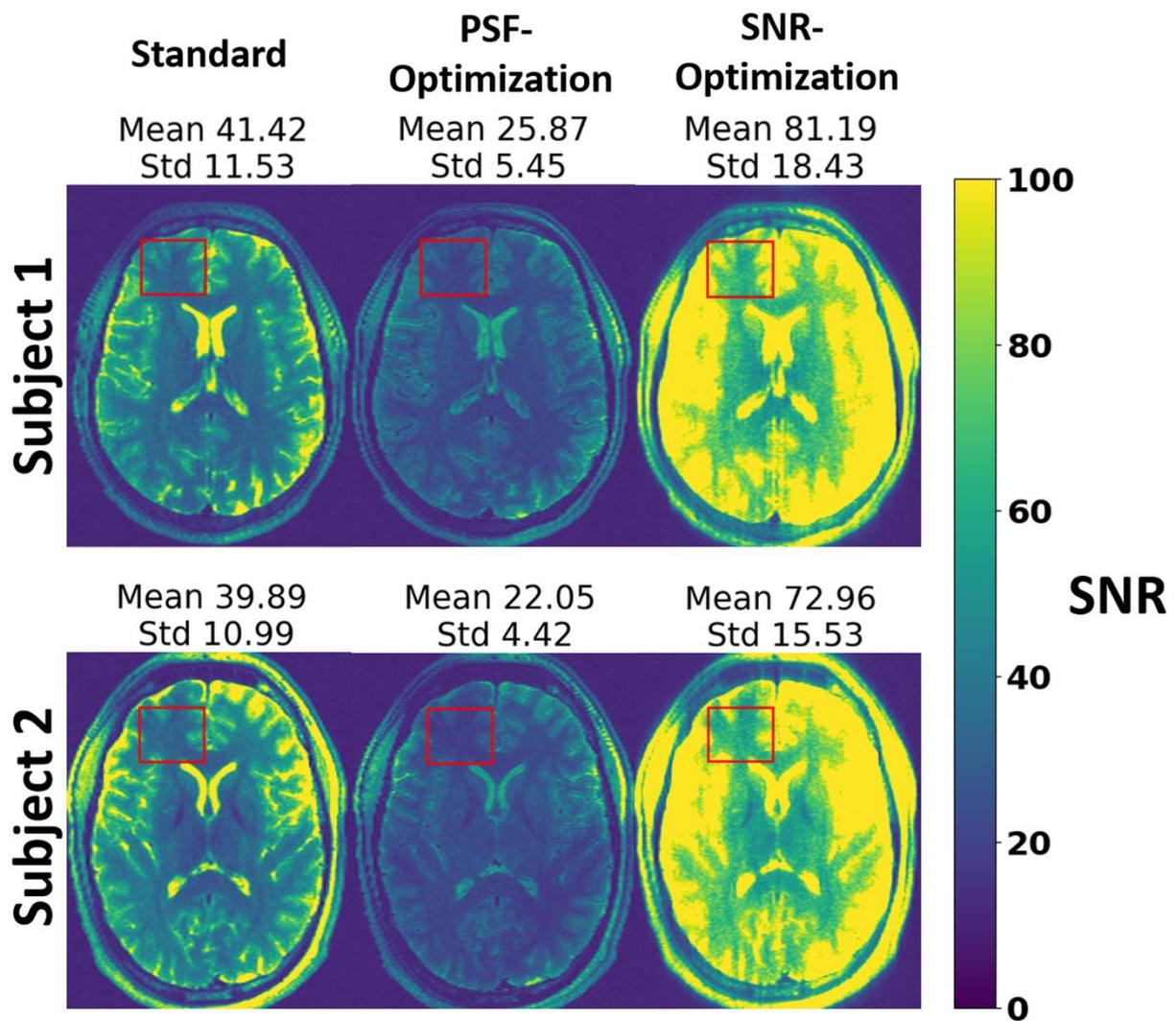

Figure 6

SNR maps of images of subject 1 and 2 using the standard, PSF- and SNR-optimized VFA schemes for the Pulseq 3D FSE sequence with T2w Monte Carlo simulations (also known as pseudo-replicas) were employed to obtain the SNR maps. Within a white and gray matter ROI, the PSF optimization yields an average SNR which is decreased by 37.4% relative to the standard for subject 1 while the SNR optimization yields an average SNR that is increased by 96.1% relative to the standard Note that similar observations were obtained for subject 2, where the PSF optimization shows decreased average SNR reduced by 44.6% relative to standard while the SNR optimization yields an average SNR elevated by 83.0% relative to the standard.

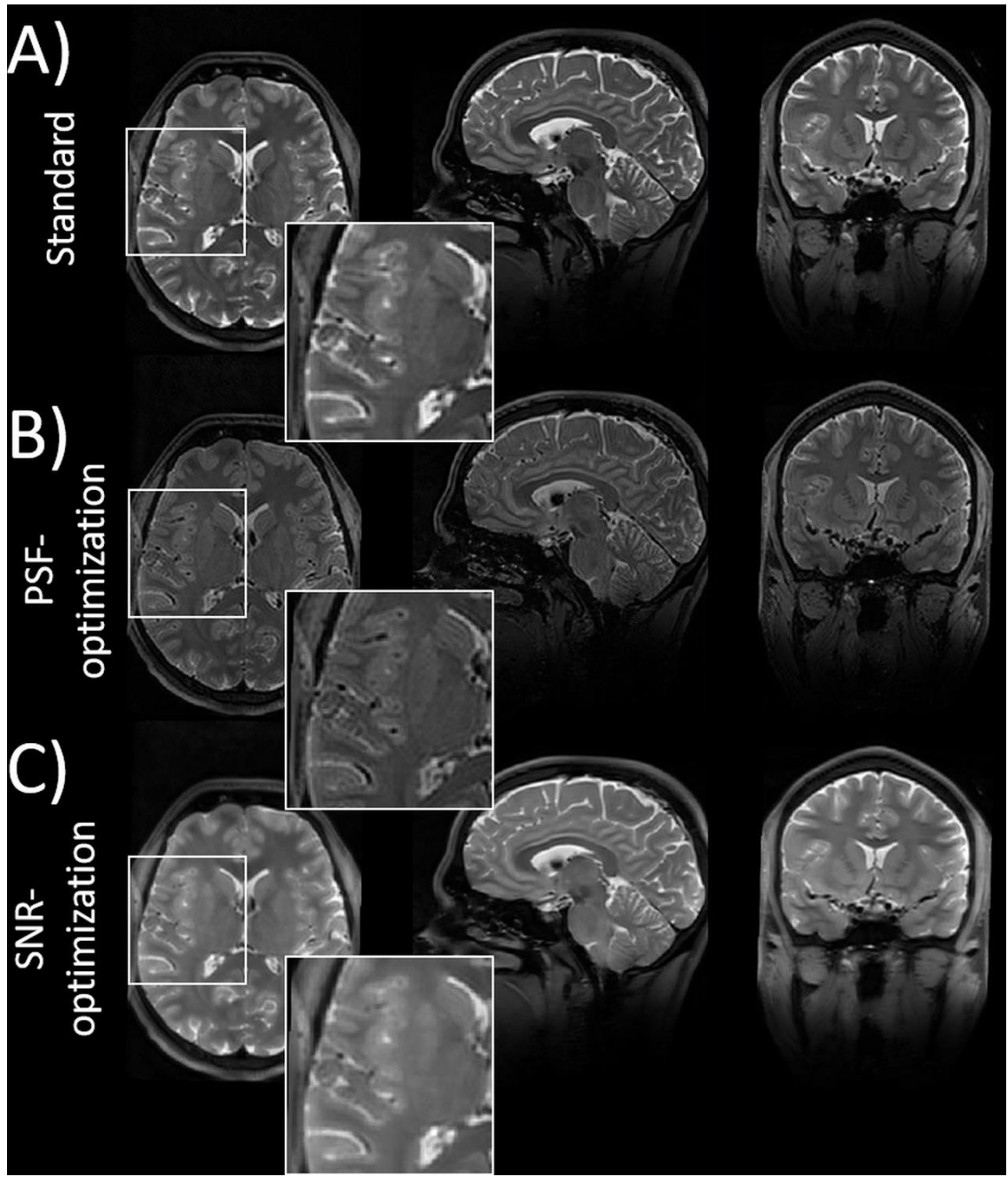

Figure 7

Images of subject 3 obtained via a vendor 3D FSE sequence with T2w using parallel transmit for standard, PSF- and SNR-optimized VFA schemes (A-C, respectively). In general, observations concluded from the Pulseq standard, are validated. The PSF optimization yields white and gray matter structures that are better resolved in comparison to the standard, especially blood vessels, while the SNR optimization provides elevated signal magnitudes with increased image blurring. Note that signal variations throughout the FOV, which are present in the Pulseq 3D FSE, are strongly minimized due to the use of parallel transmit. Note that all VFA schemes have equal relative SAR.

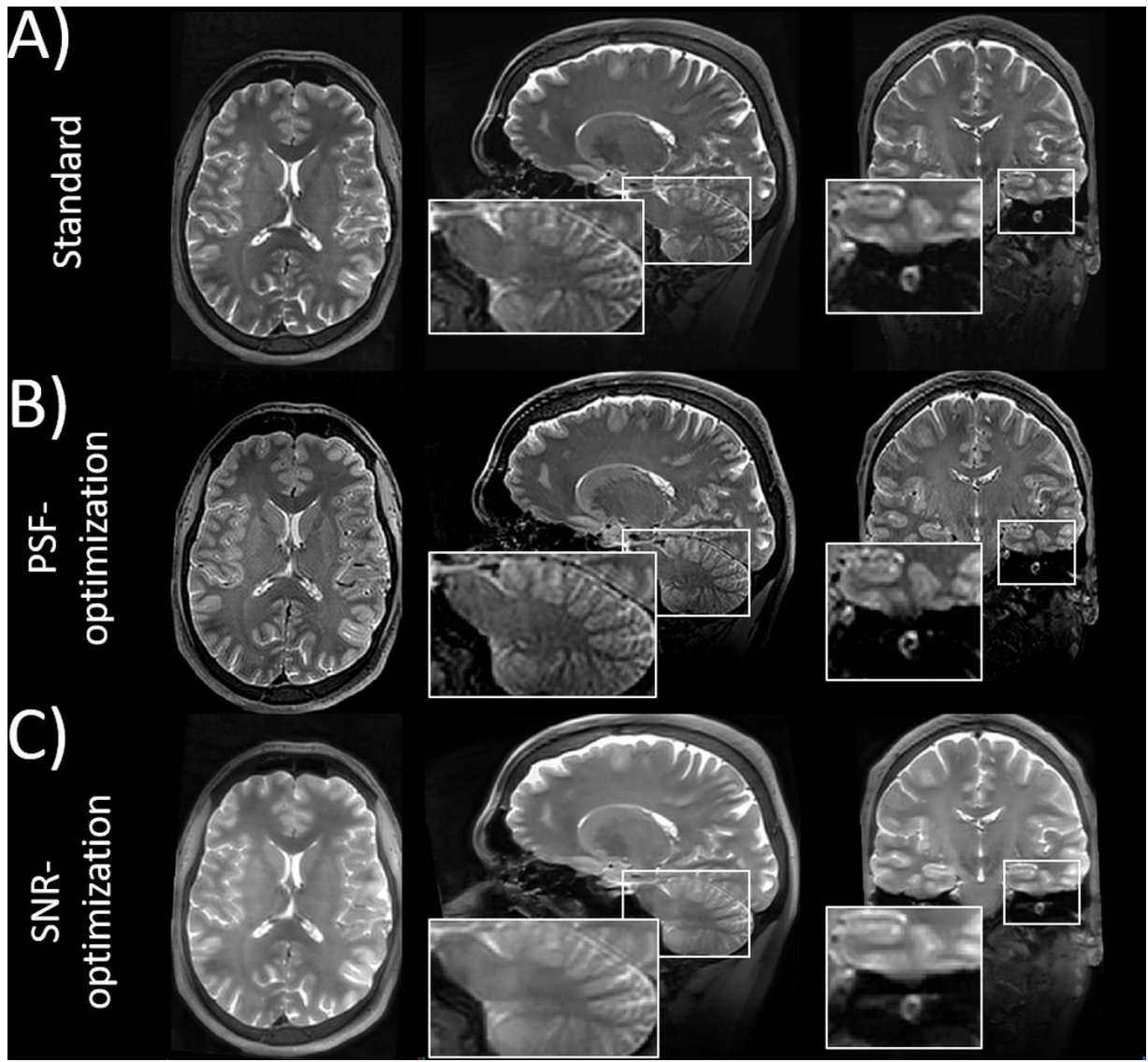

Figure 8

Images of subject 4 obtained via a vendor 3D FSE sequence with T2w using parallel transmit for standard, PSF- and SNR-optimized VFA schemes (A-C, respectively). The PSF-optimization provides better resolved structures in the cerebellum relative to the standard. Note that the blurring of the cochlea is strongly reduced, too, which is particularly helpful for segmentation tasks. On the other hand, the SNR-optimization provides strongly enhanced signal magnitudes for both the cerebellum and cochlea, however, at the cost of increased blurring.

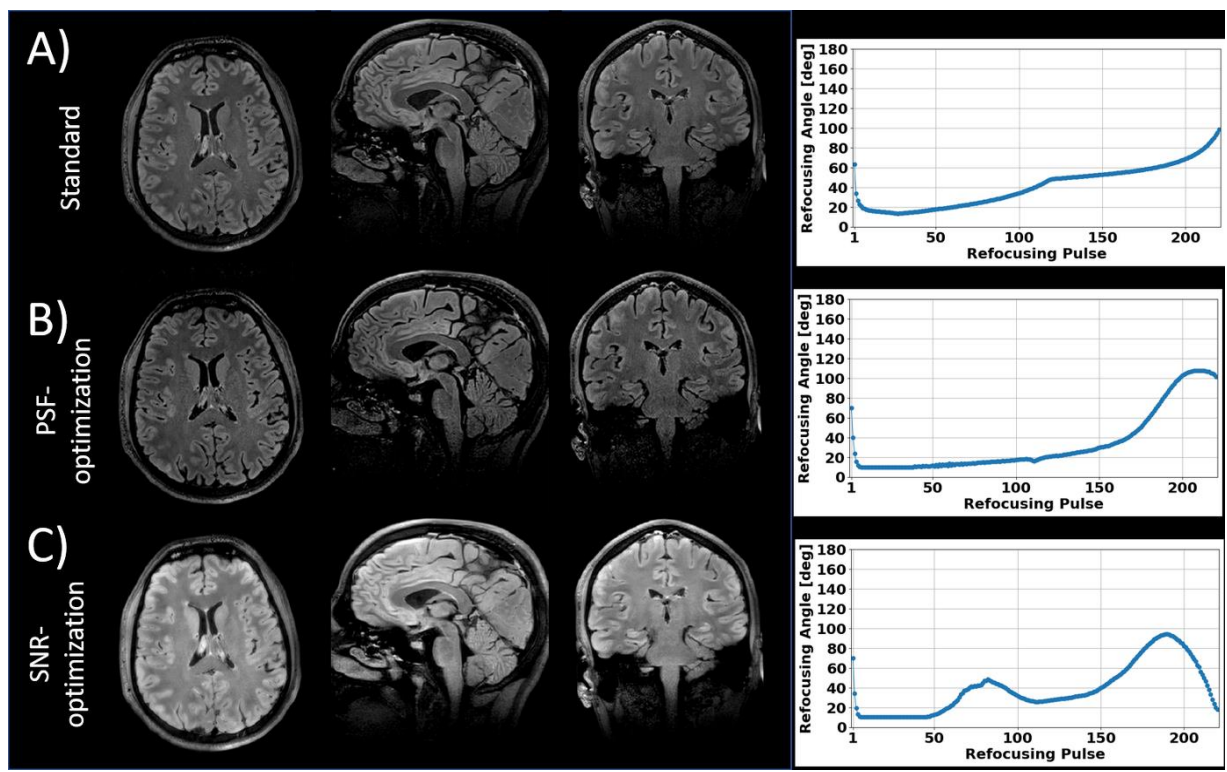

Figure 9

Images of subject 5 obtained via a vendor 3D FSE sequence with FLAIR weighting using parallel transmit for standard, PSF- and SNR-optimized VFA schemes (A-C, respectively) and corresponding VFA schemes (right hand side.) The PSF-optimized images display highly resolved WM and GM structures and reduced image blurring while providing high contrast fidelity and reasonable SNR. Note that the total echo train length is 220 in this case, resulting into an echo train duration of 888.8 ms. Highly amplified signal magnitudes are provided by the SNR-optimized VFA scheme. Note that the k-space center is assigned the 75th echo, and the SNR-optimized VFA scheme reveals elevated flip angles around this echo. Note that all VFA schemes have equal relative SAR.